\documentclass[prc,twocolumn,aps,superscriptaddress,nofootinbib,noeprint,showkeys]{revtex4-2}
\usepackage{amssymb}
\usepackage{amsmath,bm}
\usepackage{graphicx}
\usepackage[normalem]{ulem}
\usepackage{bbding}
\usepackage{booktabs}
\usepackage[usenames, dvipsnames]{xcolor}
\usepackage{lipsum}
\usepackage{multirow}
\usepackage{makecell}
\usepackage{colortbl}
\usepackage{threeparttable}

\usepackage[colorlinks,
bookmarks=true,
linkcolor=blue,
urlcolor=blue,
anchorcolor=black,
citecolor=blue
]{hyperref}

\renewcommand{\sout}{\bgroup \color{red} \ULdepth=-.5ex \ULset}

\AtBeginDocument{
\heavyrulewidth=.08em
\lightrulewidth=.05em
\cmidrulewidth=.03em
\belowrulesep=.65ex
\belowbottomsep=0pt
\aboverulesep=.4ex
\abovetopsep=0.2pt
\cmidrulesep=0.5pt
\cmidrulekern=.5em
\defaultaddspace=2.5em
}

\begin{document}

\title{Evidence for strong isovector nuclear spin-orbit interaction}

\author{Tong-Gang Yue}
\affiliation{State Key Laboratory of Dark Matter Physics, Key Laboratory for Particle Astrophysics and Cosmology (MOE), and Shanghai Key Laboratory for Particle Physics and Cosmology, School of Physics and Astronomy,
Shanghai Jiao Tong University, Shanghai 200240, China}
\affiliation{Tsung-Dao Lee Institute, Shanghai Jiao Tong University, Shanghai 201210, China}

\author{Zhen Zhang}
\thanks{Corresponding author}
\email{zhangzh275$@$mail.sysu.edu.cn}
\affiliation{Sino-French Institute of Nuclear Engineering and Technology,
Sun Yat-sen University, Zhuhai 519082, China}

\author{Lie-Wen Chen}
\thanks{Corresponding author}
\email{lwchen$@$sjtu.edu.cn}
\affiliation{State Key Laboratory of Dark Matter Physics, Key Laboratory for Particle Astrophysics and Cosmology (MOE), and Shanghai Key Laboratory for Particle Physics and Cosmology, School of Physics and Astronomy,
Shanghai Jiao Tong University, Shanghai 200240, China}

\date{\today}

\begin{abstract}
The nucleon spin-orbit interaction is a cornerstone of nuclear structure theory, yet its isospin dependence remains elusive owing to the lack of clean experimental probes. Here we show that the charge-weak form factor difference in ${}^{48}$Ca, recently extracted in a model-independent manner by the CREX experiment, exhibits strong sensitivity to the isovector spin-orbit interaction. Using Skyrme-like energy density functionals, we demonstrate that a significantly enhanced isovector spin-orbit interaction, about four times stronger than conventional parametrizations, can resolve the PREX-CREX puzzle, which has challenged modern nuclear theories and our understanding of nuclear symmetry energy, while maintaining a good description of nuclear bulk properties and well-established shell structure of finite nuclei. This enhanced isovector spin-orbit interaction also provides a novel mechanism for the emergence of the $N = 14$, $16$, $32$ and $34$ magic numbers in neutron-rich nuclei on the mean-field level.  These findings point to a strong isospin dependence of the nucleon spin-orbit interaction, which is expected to have important implications for nuclear structures, electroweak nuclear processes, and related problems in nuclear astrophysics.
\end{abstract}
\maketitle

The spin-orbit (SO) interaction, a relativistic effect, plays a crucial role across many branches of physics---from elementary quark-gluon dynamics to macroscopic phenomena in condensed matter and optics. In nuclear physics, the strong SO interaction introduced by Mayer~\cite{Mayer:1949pd} and Jensen~\cite{Haxel:1949fjd} provided the first successful explanation of nuclear magic numbers. Since then, the SO interaction has become a cornerstone of nuclear structure theory, governing shell structure, nuclear stability, and shell evolution.
Because nuclei contain both neutrons and protons, the SO interaction may, in principle, exhibit an isospin dependence, referred to as the isovector spin-orbit (IVSO) interaction. Although conventional nuclear energy-density functionals (EDFs) usually assume the IVSO interaction to be weak, its actual strength remains poorly constrained due to the lack of clean and sensitive experimental probes~\cite{Bender:2003jk}. Here we show that parity-violating electron scattering (PVES) offers a new opportunity to probe the strength of the IVSO interaction.

Recently, the PREX-II and CREX PVES experiments determined the charge-weak form-factor difference $\Delta F_{\rm CW}\equiv F_{\rm C}-F_{\rm W}$ for $^{208}$Pb and $^{48}$Ca---denoted as
$\Delta F_{\rm CW}^{208}$ and $\Delta F_{\rm CW}^{48}$, respectively---with minimal model dependence~\cite{PREX:2021umo,CREX:2022kgg}.
These observables are closely linked to the neutron-skin thickness $\Delta r_{\rm np}$ and thereby to the density dependence of the symmetry energy~\cite{Brown:2000pd}, and have been therefore expected to impose strong constraints on both the slope parameter $L$ of the symmetry energy and $\Delta r_{\rm np}$.
Unexpectedly, however, modern EDFs fail to reproduce $\Delta F_{\rm CW}^{208}$ and $\Delta F_{\rm CW}^{48}$ simultaneously within $1\sigma$~\cite{CREX:2022kgg}.
This striking inconsistency---the PREX-CREX puzzle---indicates that a key piece of physics is missing from current EDF frameworks.

Here we show that this missing piece is a strong IVSO interaction.
Using extended Skyrme EDFs, we find that $\Delta F_{\rm CW}^{48}$ is remarkably sensitive to the IVSO interaction, and an IVSO strength about four times larger than in conventional parametrizations resolves the PREX-CREX puzzle while preserving good description for nuclear global properties. The same strong IVSO interaction naturally generates the new magic numbers $N=14$, $16$, $32$, and $34$ in neutron-rich oxygen and calcium isotopes.
Together, these findings provide compelling evidence for a strong IVSO interaction in nuclei, with broad implications for nuclear structure, electroweak probes, and astrophysics.

We employ a nonrelativistic EDF based on the extended Skyrme interaction~\cite{Zhang:2015vaa}, supplemented by a zero-range tensor force and a density-dependent pairing interaction.
Within this EDF, the SO contribution to the nuclear binding energy is
\begin{equation}
E_{\rm so} = \int {\rm d}^3r \left[ \frac{b_{\rm IS}}{2}\,\bm{J}\cdot \bm{\nabla}\rho
+ \frac{b_{\rm IV}}{2}\,(\bm{J}_{\rm n}-\bm{J}_{\rm p})\cdot\bm{\nabla}(\rho_{\rm n}-\rho_{\rm p}) \right],
\label{Eq:eso_short}
\end{equation}
where $\rho=\rho_{\rm n}+\rho_{\rm p}$ and $\bm{J}=\bm{J}_{\rm n}+\bm{J}_{\rm p}$ are the isoscalar nucleon density and SO density, and $\rho_{\rm n(p)}$ and $\bm{J}_{\rm n(p)}$ denote neutron (proton) densities and SO densities.
The parameters $b_{\rm IS}$ and $b_{\rm IV}$ quantify the isoscalar and isovector SO strengths.
Whereas conventional Skyrme EDFs impose $b_{\rm IV}=b_{\rm IS}/3$, we treat
$b_{\rm IV}$ as a free parameter.

New EDFs are constructed by fitting to ground-state observables of doubly-magic nuclei (binding energies, charge radii, diffraction radii, and surface thicknesses), the isoscalar giant monopole resonance in $^{208}$Pb, and selected constraints on nuclear matter from heavy-ion collisions and microscopic calculations.
The pairing strength is adjusted to reproduce the empirical pairing gap in $^{120}$Sn in Hartree-Fock-Bogoliubov calculations~\cite{Bender:2003jk}.
Full details of the functional and the fit protocol are provided in the Supplementary material.

To isolate IVSO effects, we focus on two representative EDFs:
eS53, with $b_{\rm IV} = 53~\mathrm{MeV\,fm}^5\simeq b_{\rm IS}/3$, corresponding to a conventional weak IVSO interaction; and
eS250, with $b_{\rm IV}=250~\mathrm{MeV\,fm}^5$, representing a strong IVSO interaction.
All other parameters of eS53 and eS250 are identical, ensuring that the differences in their predictions can be attributed solely to $b_{\rm IV}$.

\begin{figure}[t]
    \centering
    \includegraphics[width=0.9\linewidth]{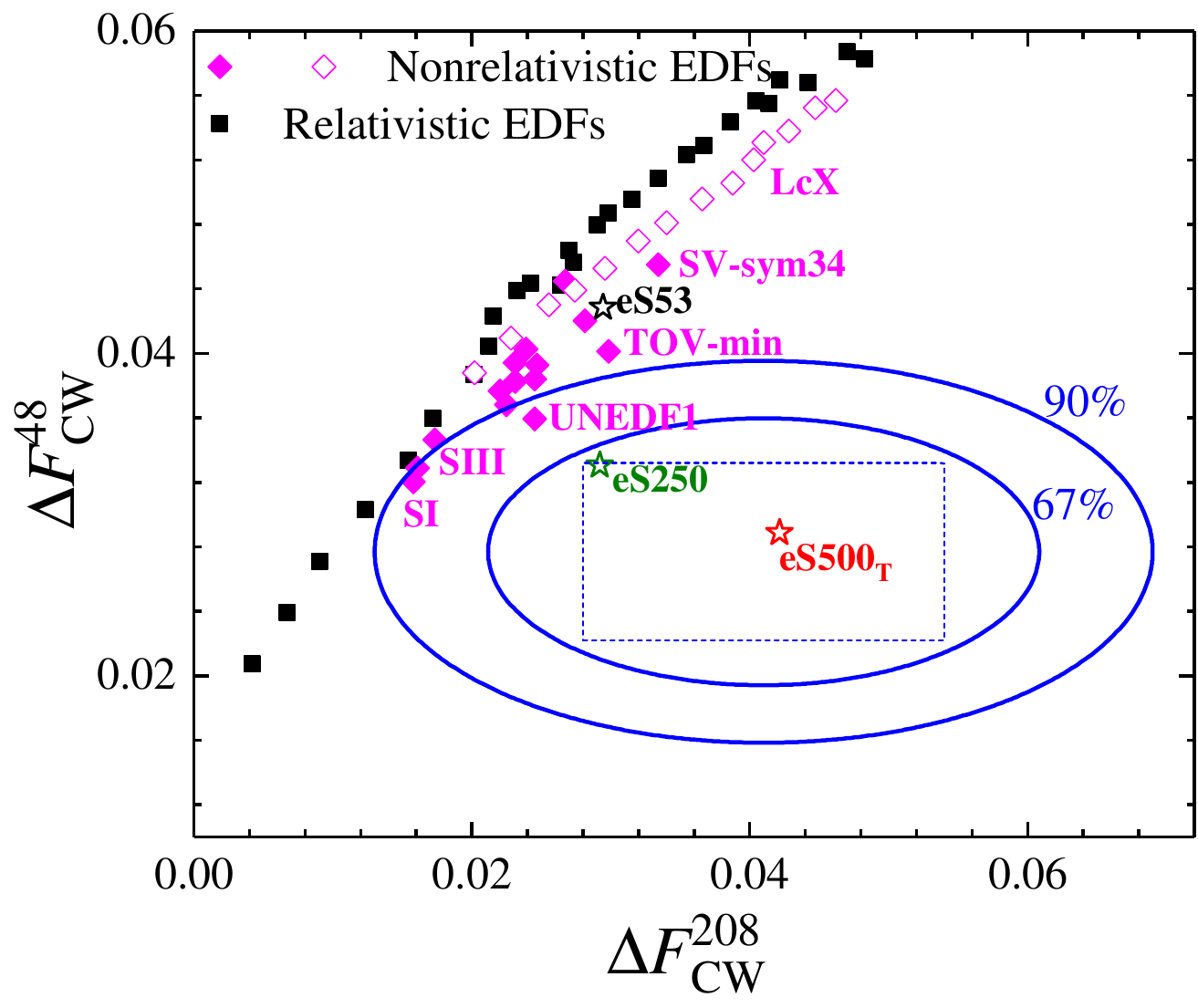}
    \caption{Charge-weak form-factor differences in $^{208}$Pb and $^{48}$Ca.
    Predictions from the three Skyrme-like EDFs eS53, eS250, and eS500$_{\rm T}$ constructed in this work are shown as open stars.
    Also included are predictions from various covariant (squares) and nonrelativistic (diamonds) EDFs~\cite{CREX:2022kgg,Yue:2021yfx}.
    The ellipses depict the joint PREX-II and CREX $67\%$ and $90\%$ probability contours, while the rectangle indicates the marginal $68.3\%$ confidence intervals for $\Delta F_{\rm CW}^{48}$ and $\Delta F_{\rm CW}^{208}$~\cite{CREX:2022kgg}.}
    \label{Fig:DFCW}
\end{figure}

The PREX-II and CREX measurements report
$\Delta F_{\rm CW}^{208}(q=0.3977~\mathrm{fm}^{-1})=0.041\pm0.013$
and
$\Delta F_{\rm CW}^{48}(q=0.8733~\mathrm{fm}^{-1})=0.0277\pm0.0055$, respectively~\cite{PREX:2021umo,CREX:2022kgg}. Fig.~\ref{Fig:DFCW} compares these values with predictions from a large set of relativistic and nonrelativistic EDFs~\cite{CREX:2022kgg,Yue:2021yfx}, together with eS53, eS250, and a more extreme strong-IVSO EDF eS500$_{\rm T}$ with $b_{\rm IV}=500~\mathrm{MeV~fm}^5$.
One sees the conventional EDFs (including eS53) form a correlation band that systematically deviates from the joint 1$\sigma$ confidence ellipse of the PREX-II and CREX measurements.
This systematic deviation is the PREX-CREX puzzle~\cite{CREX:2022kgg}.

The eS250 EDF with a strong IVSO interaction behaves very differently.
Compared with eS53, it lowers the predicted value of $\Delta F_{\rm CW}^{48}$ while leaving $\Delta F_{\rm CW}^{208}$ essentially unchanged, bringing both into agreement with the joint PREX-II and CREX $1\sigma$ region.
At the same time, eS250 maintains a good description of nuclear masses, charge radii, SO splittings, and dipole polarizabilities in $^{48}$Ca and $^{208}$Pb, and remains compatible with microscopic neutron-matter calculations (see Tables S1-S3 (online) and Figs. S1-S3 (online)).
The even stronger-IVSO functional eS500$_{\rm T}$ can further improve the agreement with the PREX-II and CREX data but tends to overpredict dipole polarizabilities and the stiffness of neutron matter (see Table S3 (online) and
Fig. S3 (online)), and is therefore disfavored. It is worth noting that the eS250 predicts a relatively larger $\Delta F_{\rm CW}^{208}$ compared to the mean value of ab initio predictions obtained with 34 non-implausible chiral interactions reported in Ref.~\cite{Hu:2021trw}, while remaining within the spread of these interactions.
Such a difference points to the importance of uncertainty quantification on both the ab initio and EDF sides.

A correlation analysis (see Fig. S4 (online)) shows that
$\Delta F_{\rm CW}^{208}$ is mainly controlled by the symmetry-energy slope $L$, whereas $\Delta F_{\rm CW}^{48}$ depends strongly on both $L$ and $b_{\rm IV}$.
In particular, $\Delta F_{\rm CW}^{48}$ exhibits a pronounced negative correlation with $b_{\rm IV}$, a feature that is essential for reconciling the PREX-II and CREX results.
We note that combining the $\Delta F_{\rm CW}$,  dipole polarizabilities, and neutron-matter EOS yields
$L\approx 58$~MeV and $b_{\rm IV}\approx 250~\mathrm{MeV\,fm}^5$.

The different roles played by $b_{\rm IV}$ in $^{208}$Pb and $^{48}$Ca can be understood from the isovector SO density $\bm{J}_{\rm n}-\bm{J}_{\rm p}$ entering Eq.~(\ref{Eq:eso_short}).
For spherical nuclei, the SO density can be written as~\cite{Colo:2007cwc}
\begin{eqnarray}
\bm{J}_{q}(r) &=& \frac{1}{4 {\rm \pi} r^3} \sum_i v_i^2\left(2 j_i+1\right) \notag \\
 &&\times\left[j_i\left(j_i+1\right)-l_i\left(l_i+1\right)-\frac{3}{4}\right] R_i^2(r),
\label{Eq:Jq}
\end{eqnarray}
where the sum runs over all orbitals of isospin $q$, $v_i^2$ is the occupation probability, and $R_i(r)$ is the radial wave function.
States with $j_{>}=l+1/2$ and $j_{<}=l-1/2$, which form a pair of SO partners, contribute with opposite signs. As a result, $\bm{J}_{q}$ is small when SO partners are both filled (SO-saturated case), and becomes large when only one partner is occupied.

\begin{figure*}[t]
    \centering
    \includegraphics[width=0.8\linewidth]{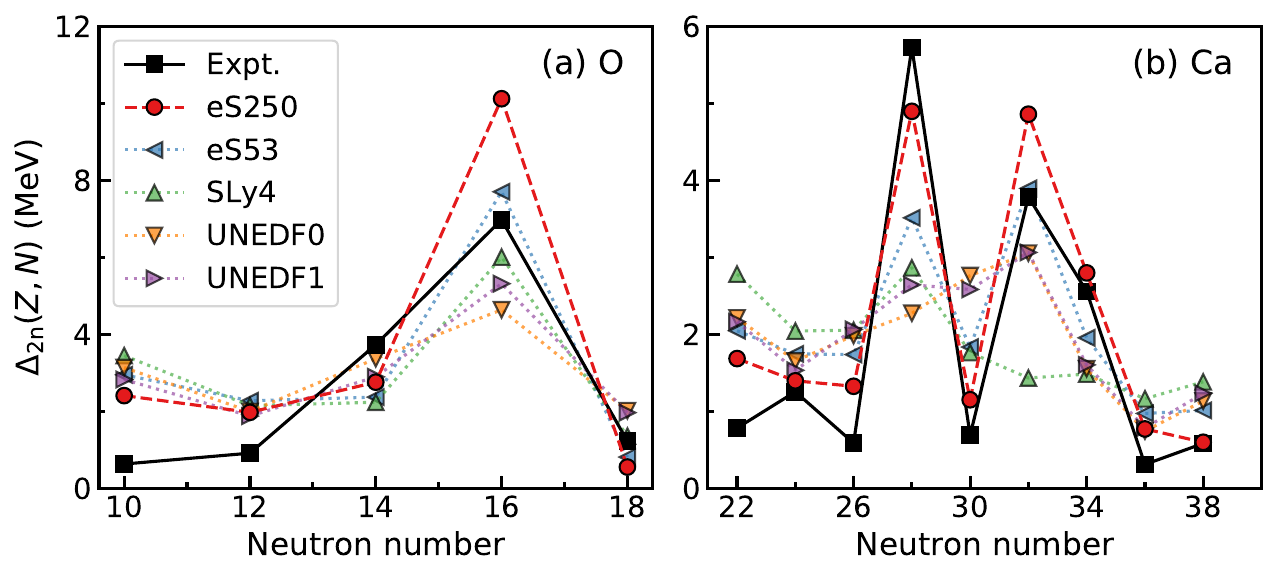}
    \caption{Two-neutron shell gaps $\Delta_{2\rm{n}}$ for O (a) and Ca (b) isotopes.
    Predictions of eS250 and eS53 are compared with experimental data~\cite{Wang:2021xhn} and with the Skyrme EDFs SLy4~\cite{Chabanat:1997un}, UNEDF0~\cite{Kortelainen:2010hv}, and UNEDF1~\cite{Kortelainen:2011ft}.}
    \label{Fig:dltS2n}
\end{figure*}

In $^{208}$Pb, the SO-unsaturated proton and neutron orbitals ($1\mathrm{h}_{\rm 11/2}$ and $1\mathrm{i}_{\rm 13/2}$) generate comparable positive contributions to $\bm{J}_{\rm p}$ and $\bm{J}_{\rm n}$, resulting in a small
$\bm{J}_{\rm n}-\bm{J}_{\rm p}$ and consequently a weak IVSO effect on $\Delta F_{\rm CW}^{208}$.
In $^{48}$Ca, however, all proton SO partners are filled, while the neutron $1\mathrm{f}_{\rm 7/2}$ shell is fully occupied and its SO partner $1\mathrm{f}_{\rm 5/2}$ is empty.
This configuration produces a large
$\bm{J}_{\rm n}-\bm{J}_{\rm p}$, and the IVSO term proportional to
$b_{\rm IV}(\bm{J}_{\rm n}-\bm{J}_{\rm p})\cdot\nabla(\rho_{\rm n}-\rho_{\rm p})$ modifies the central mean field and
induces a global redistribution of neutron and proton densities.
These density rearrangements dominate
over the direct changes in the nucleon SO potential (see Fig. S5 (online)), explaining why
the $\Delta F_{\rm CW}^{48}$---but
not the $\Delta F_{\rm CW}^{208}$---is highly sensitive to $b_{\rm IV}$.

A strong IVSO interaction also leaves a clear imprint on shell evolution.
Fig.~\ref{Fig:dltS2n} shows the two-neutron shell gaps
$\Delta_{\rm 2n}(Z,N)\equiv S_{\rm 2n}(Z,N)-S_{\rm 2n}(Z,N+2)$
for oxygen and calcium isotopes, where $S_{\rm 2n}(Z,N)$ denotes the two-neutron separation energy. Results for eS53 and eS250 are compared with experimental data~\cite{Wang:2021xhn} and with conventional Skyrme EDFs SLy4~\cite{Chabanat:1997un}, UNEDF0~\cite{Kortelainen:2010hv}, and UNEDF1~\cite{Kortelainen:2011ft}.
The weak-IVSO eS53 EDF and conventional parametrizations predict only modest structures at
$N=14$, $16$, $32$, and $34$, underestimating the magic or semi-magic behavior inferred from separation
energies, radii, and spectroscopy~\cite{Ye:2024slx}. In contrast, the strong-IVSO eS250 functional
produces a pronounced kink at $N=16$ (and a weaker one at $N=14$) in oxygen and sizable gaps at
$N=32$ and $34$ in calcium, while preserving the traditional $N=28$ closure.

The underlying mechanism is the enhancement of SO splittings driven by a large $b_{\rm IV}$. In
$^{22,24}$O, the six neutrons occupying the $1\mathrm{d}_{\rm 5/2}$ orbital generate a sizable
$\bm{J}_{\rm n}-\bm{J}_{\rm p}$, and a strong IVSO interaction further increases the
$1\mathrm{d}_{\rm 5/2}$-$1\mathrm{d}_{\rm 3/2}$ splitting with
enlarging gaps below and above the $2\mathrm{s}_{\rm 1/2}$ level
(which lies between $1\mathrm{d}_{\rm 5/2}$ and $1\mathrm{d}_{\rm 3/2}$), stabilizing the
$N=14$ and $N=16$ closures. In the calcium isotopes, the strengthened SO splittings between the $2\mathrm{p}_{\rm 3/2}$
and $1\mathrm{f}_{\rm 5/2}$ orbitals lead to substantial gaps between the $2\mathrm{p}_{\rm 3/2}$ and $2\mathrm{p}_{\rm 1/2}$ states in $^{52}$Ca
and between $2\mathrm{p}_{\rm 3/2}$ and $1\mathrm{f}_{\rm 5/2}$ in $^{54}$Ca, generating the new magic numbers $N=32$ and
$N=34$ (see Fig. S6 (online)). This provides a mean-field mechanism, complementary to tensor-force
effects~\cite{Otsuka:2018bqq}, for the emergence of these new magic numbers.

Our results therefore suggest a strong IVSO interaction with the strength $b_{\rm IV}\sim 250~\mathrm{MeV\,fm}^5$, about four times larger than in conventional Skyrme EDFs, as the value preferred by PREX-II and CREX together with other nuclear-structure and neutron-matter constraints.
The associated symmetry-energy slope $L\approx 58$~MeV (see the eS250 prediction in Table S2 (online)) is consistent with the current world average value $L=58.7\pm28.1$ MeV~\cite{Oertel:2016bki}.
%global analyses of terrestrial experiments and astrophysical observations
Given the central role of the symmetry energy in governing the pressure of neutron-rich matter, this inference also carries important implications for core-collapse supernovae, neutron-star structure, and neutron-star mergers. Improved knowledge
of neutron and weak-charge density distributions in nuclei further refines the nuclear inputs for weak-interaction processes
such as coherent elastic neutrino-nucleus scattering and dark-matter searches~\cite{Zheng:2014nga, Piekarewicz:2025lel}.

It is interesting to mention that, in relativistic density functional theory, the SO interaction naturally arises from the Lorentz structure of the nucleon mean fields. From this perspective, the enhanced IVSO interaction identified here motivates further exploration of isovector degrees of freedom---such as isovector-scalar and isovector-tensor couplings---within covariant density functional frameworks.

Furthermore, the strong IVSO interaction inferred here is expected to influence a wide range of phenomena, including the neutron drip line, the $r$-process path, and the stability of superheavy nuclei.
Future high-precision PVES experiments, such as the Mainz Radius Experiment~\cite{Schlimme:2024eky}, with targets that have distinct SO shell structures (e.g., $^{208}$Pb, $^{62}$Ni, $^{48}$Ca, and $^{90}$Zr), will be crucial for pinning down the IVSO strength and its interplay with the symmetry energy.

\section*{Conflict of interest}
The authors declare that they have no conflict of interest.

\section*{Acknowledgements}
This work was supported in part by the National Natural Science Foundation of China (12235010, 12575137 and 11905302), the National SKA Program of China (2020SKA0120300), and the Science and Technology
Commission of Shanghai Municipality (23JC1402700).

\section*{Data availability}
All the data supporting the findings in this work are available within the manuscript and any additional data are available from the corresponding authors upon reasonable request.

\section*{Author Contributions}
Tong-Gang Yue performed the theoretical calculations. Zhen Zhang and Lie-Wen Chen supervised the theoretical calculations. All authors jointly analyzed the data, and contributed to the interpretation of the results and to the writing of the manuscript.

\bibliography{RefMain}

\end{document}

%\preprint{APS/123-QED}

\title{Supplementary Material for ``Evidence for strong isovector nuclear spin-orbit interaction''}

\author{Tong-Gang Yue}
\affiliation{State Key Laboratory of Dark Matter Physics, Key Laboratory for Particle Astrophysics and Cosmology (MOE), and Shanghai Key Laboratory for Particle Physics and Cosmology, School of Physics and Astronomy,
Shanghai Jiao Tong University, Shanghai 200240, China}
\affiliation{Tsung-Dao Lee Institute, Shanghai Jiao Tong University, Shanghai 201210, China}

\author{Zhen Zhang}
\thanks{Corresponding author}
\email{zhangzh275$@$mail.sysu.edu.cn}
\affiliation{Sino-French Institute of Nuclear Engineering and Technology,
Sun Yat-sen University, Zhuhai 519082, China}

\author{Lie-Wen Chen}
\thanks{Corresponding author}
\email{lwchen$@$sjtu.edu.cn}
\affiliation{State Key Laboratory of Dark Matter Physics, Key Laboratory for Particle Astrophysics and Cosmology (MOE), and Shanghai Key Laboratory for Particle Physics and Cosmology, School of Physics and Astronomy,
Shanghai Jiao Tong University, Shanghai 200240, China}

\date{\today}% It is always \today, today,

\begin{abstract}
 This Supplementary Material provides additional details and includes three tables and six figures. Section~\ref{sec:smsubsec1} presents the formalism of the extended Skyrme energy density functional. Section~\ref{sec:smsubsec2} describes the computation of charge and weak form factors. Finally, section~\ref{sec:smsubsec3} details the fitting procedure used to construct the new energy density functionals..
\end{abstract}

%\pacs{21.65.Ef, 21.65.Cd, 24.30.Cz, 21.30.Fe }% PACS, the Physics and Astronomy
% \keywords{isovector spin-orbit interaction, formfactor, magic number}%Use showkeys class option if keyword
%                               %display desired

\maketitle

\section{Extended Skyrme energy density functional}\label{sec:smsubsec1}

In the present work, we use a nonrelativistic nuclear energy density functional based on the extended Skyrme interaction~\cite{Chamel:2009yx}, with further incorporation of a zero-range tensor force~\cite{Stancu:1977va}. The adopted extended Skyrme interaction $v_{\rm{eSky}}$ differs from conventional Skyrme interaction $v_{\rm{Sky}}$ by two additional terms, that is
\begin{eqnarray}
\begin{aligned}
v_{\rm{eSky}}=v_{\rm{Sky}}+v^{\prime},
\end{aligned}
\end{eqnarray}
where the conventional Skyrme interaction $v_{\rm{Sky}}$ reads (see, e.g., Ref.~\cite{Chabanat:1997qh}):
\begin{eqnarray}\
\begin{aligned}
& v_{\rm Sky}\left(\bm{r}_1, \bm{r}_2\right)=t_0\left(1+x_0 P_{\rm \sigma}\right) \delta(\mathbf{r}) +\frac{1}{2} t_1\left(1+x_1 P_{\rm \sigma}\right)\left[\mathbf{k}^{\prime 2} \delta(\mathbf{r})+\delta(\mathbf{r}) \mathbf{k}^2\right] +t_2\left(1+x_2 P_{\rm \sigma}\right) \mathbf{k}^{\prime} \cdot \delta(\mathbf{r}) \mathbf{k} \\
&+\frac{1}{6} t_3\left(1+x_3 P_{\sigma}\right)[\rho(\mathbf{R})]^\alpha \delta(\mathbf{r}) + i W_0 (\bm{\sigma}_1+\bm{\sigma}_2)\cdot\left[\mathbf{k}^{\prime} \times \delta(\mathbf{r}) \mathbf{k}\right],
\end{aligned}
\end{eqnarray}
and the two additional terms are represented by
\begin{equation}
\begin{aligned}
v^{\prime} =\frac{1}{2} t_4\left(1+x_4 P_{\rm \sigma}\right)\left[\mathbf{k}^{\prime 2} \rho(\bm{R})^\beta \delta(\mathbf{r})+\delta(\mathbf{r}) \rho(\bm{R})^\beta \mathbf{k}^2\right] + t_5\left(1+x_5 P_{\rm \sigma}\right) \mathbf{k}^{\prime} \cdot \rho(\bm{R})^\gamma \delta(\mathbf{r}) \mathbf{k}.
\end{aligned}
\end{equation}
Here, $ \bm{r}=\bm{r}_1-\bm{r}_2$, $ \bm{R}=\frac{1}{2}\left(\bm{r}_1+\bm{r}_2\right)$, $\bm{k}=\frac{1}{2 i}\left(\bm{\nabla}_1-\bm{\nabla}_2\right)$ is the relative momentum with $\bm{k}^{\prime}$ being its conjugate acting on the left, the $\bm{\sigma}$ are Pauli spin operators, $P_{\rm \sigma}=\frac{1}{2}\left(1+\bm{\sigma}_1 \cdot \bm{\sigma}_2\right)$ is the spin exchange operator,
and $\rho(\bm{R})=\rho_{\rm n}(\bm{R})+\rho_{\rm p}(\bm{R})$ is the total local density with $\rho_{\rm n}$ and $\rho_{\rm p}$ being neutron and proton densities, respectively. The $t_0\sim t_5$, $x_0\sim x_5$, $\alpha$, $\beta$ and $\gamma$ are adjustable ``Skyrme parameters'', and $W_0$ is the spin-orbit (SO) interaction parameter. For simplicity, both $\beta$ and $\gamma$ are set to be unity as in Ref.~\cite{Zhang:2015vaa}.

For the tensor force, we use the following zero-range form~\cite{Stancu:1977va}
\begin{equation}
\begin{aligned}
V_T= & \frac{1}{2} T\left\{\left[\left(\bm{\sigma}_1 \cdot \mathbf{k}^{\prime}\right)\left(\bm{\sigma}_2 \cdot \mathbf{k}^{\prime}\right)-\frac{1}{3} \mathbf{k}^{\prime 2}\left(\bm{\sigma}_1 \cdot \bm{\sigma}_2\right)\right] \delta\left(\mathbf{r}\right) + \delta\left(\mathbf{r}\right)\left[\left(\bm{\sigma}_1 \cdot \mathbf{k}\right)\left(\bm{\sigma}_2 \cdot \mathbf{k}\right)-\frac{1}{3} \mathbf{k}^2\left(\bm{\sigma}_1 \cdot \bm{\sigma}_2\right)\right]\right\} \\
& +U\left\{\left(\bm{\sigma}_1 \cdot \mathbf{k}^{\prime}\right) \delta\left(\mathbf{r}\right)\left(\bm{\sigma}_2 \cdot \mathbf{k}\right) - \frac{1}{3}\left(\bm{\sigma}_1 \cdot \bm{\sigma}_2\right)\left[\mathbf{k}^{\prime} \cdot \delta\left(\mathbf{r}\right) \mathbf{k}\right]\right\} ,
\end{aligned}
\end{equation}
where $T$ and $U$ quantify the magnitude of the tensor force in states of even and odd relative motion, respectively. In practical calculations, the parameters $\alpha_{\rm T}$ and $\beta_{\rm T}$ are employed in place of $U$ and $T$, where $\alpha_{\rm T}$ and $\beta_{\rm T}$ are expressed as:
\begin{equation}
\alpha_{\rm T} = \frac{5}{12} U, \quad \beta_{\rm T} = \frac{5}{24} (T + U).
\end{equation}

In addition, for calculations of open-shell nuclei, we take into account pairing force of the form
\begin{equation}
V_{\rm p}(r,r^{\prime}) =  V_0 \left[ 1 - \frac{\rho(r)}{2\rho_0} \right] \delta(r-r^{\prime}),
\end{equation}
where $V_0$ is the pairing strength and $\rho_0 = 0.16~\rm fm^{-3}$.

Based on above extended Skyrme interaction, tensor force and pairing force, we can obtain the energy density functional within the Hartree-Fock(-Bogoliubov) (HF(B)) theory. Under the assumption of time-reversal invariance, the energy of a nucleus can be written
as the integral of a purely local energy-density functional $\mathcal{E}$. For details on the HF(B) calculation based on Skyrme EDF, see e.g. Ref.~\cite{Stoitsov:2004pe}. In the following, we focus on potential energy density $\mathcal{E}_{\rm{eSky}}$ for extended Skyrme interaction and tensor force, which can be expressed in terms of
local density $\rho_{ q}$, kinetic-energy density $\tau_{q}$ and SO density $\bm{J}_{q}$ of neutrons ($q=\rm{n}$) and protons ($q=\rm{p}$) defined by
\begin{eqnarray}
        \rho_{q}(\bm{r}) & =&\sum_{i} v_i^2\left|\varphi_i(\bm{r})\right|^2, \\
        \tau_{q}(\bm{r}) & =&\sum_i v_i^2\left|\nabla \varphi_i\bm{r})\right|^2, \\
        \bm{J}_{q}(\bm{r}) & =&-\mathrm{i} \sum_i v_i^2 \varphi_i^{+}(\bm{r}) \nabla \times \hat{\sigma} \varphi_i(\bm{r}).
\end{eqnarray}
Here, $\varphi_i$ is the wave function of canonical state $i$, $v_i$ is corresponding occupation  probability, and the sum runs over all states for the same species $q$. By further defining the isoscalar densities
\begin{eqnarray}
 \rho =\rho_{\rm n}+\rho_{\rm p},~~\tau = \tau_{\rm n}+\tau_{\rm p}, ~~\bm{J}= \bm{J}_{\rm n}+\bm{J}_{\rm p},
\end{eqnarray}
and isovector densities
\begin{eqnarray}
 \tilde{\rho} =\rho_{\rm n}-\rho_{\rm p},~~\tilde{\tau} = \tau_{\rm n}-\tau_{\rm p}, ~~\tilde{\bm{J}}= \bm{J}_{\rm n}-\bm{J}_{\rm p},
\end{eqnarray}
the $\mathcal{E}_{\rm{eSky}}$ can be written as
\begin{eqnarray} \label{Eq:Sky-HF}
    \mathcal{E}_{\rm eSky} &=&
\frac{B_0+B_3 \rho^\alpha}{2} \rho^2-\frac{B_0^{\prime}+B_3^{\prime} \rho^\alpha}{2} \tilde{\rho}^2+(B_1+B_4\rho^{\beta}+B_5\rho^{\gamma})\rho \tau -(B_1^{\prime}+B_4^{\prime}\rho^{\beta}+B_5^{\prime}\rho^{\gamma}) \tilde{\rho} \tilde{\tau}  \notag \\
& & +\frac{2B_2+(2\beta+3)B_4\rho^{\beta}-B_5\rho^{\gamma}}{4} (\nabla \rho)^2 -\frac{2B_2^{\prime}+3B_4^{\prime}\rho^{\beta}-B_5^{\prime}\rho^{\gamma}}{4}(\nabla{\tilde{\rho}})^2-\frac{\beta B_4^{\prime}}{2}\rho^{\beta-1}\tilde{\rho}\nabla \rho \cdot \nabla \tilde{\rho}
\notag \\
& & +\frac{C_1+C_2\rho^{\beta}+C_3\rho^{\gamma}}{2} \bm{J}^2 + \frac{C_1^{\prime}+C_2^{\prime} \rho^{\beta} + C_3^{\prime} \rho^{\gamma} }{2} \bm{\tilde{J}}^2  \notag \\
        & & +\frac{b_{\rm{IS}}}{2} \nabla \rho \cdot \mathbf{J} + \frac{b_{\rm{IV}}}{2} \nabla \tilde{\rho} \cdot \mathbf{\tilde{J}}
        + \frac{\alpha_T + \beta_T}{4} \bm{J}^2 + \frac{\alpha_T - \beta_T}{4} \bm{\tilde{J}}^2.
    \end{eqnarray}

Here, $b_{\rm{IS}}$ and $b_{\rm{IV}}$ represent the strengths of the isoscalar and isovector spin-orbit (IVSO) interactions, and the coefficients $B_i$, $B_i^{\prime}$, $C_i$ and $C_i^{\prime}$ are uniquely related to the Skyrme parameters $t_i$ and $x_i$ by
    \begin{equation}\label{Eq:BC}
        \begin{aligned}
        B_0&=\frac{3}{4}t_0,&B_0^{\prime}&=\frac{1}{2}t_0 (\frac{1}{2} + x_0), \\
        B_1&=\frac{3}{16} t_1+\frac{5}{16} t_2+\frac{1}{4} t_2 x_2, &B_1^{\prime}&=\frac{1}{8}\left[t_1\left(\frac{1}{2}+x_1\right)-t_2\left(\frac{1}{2}+x_2\right)\right], \\
        B_2&=\frac{9}{32} t_1-\frac{5}{32} t_2-\frac{1}{8} t_2 x_2, &B_2^{\prime}&=\frac{1}{16}\left[3 t_1\left(\frac{1}{2}+x_1\right)+t_2\left(\frac{1}{2}+x_2\right)\right], \\
        B_3&=\frac{1}{8} t_3 , &B_3^{\prime}&=\frac{1}{12} t_3 (\frac{1}{2} + x_3), \\
        B_4&=\frac{3}{16} t_4, &B_4^{\prime}&=\frac{1}{8}t_4\left(\frac{1}{2}+x_4\right), \\
        B_5&=\frac{5}{16} t_5+\frac{1}{4} t_5 x_5, &B_5^{\prime}&=-\frac{1}{8}t_5\left(\frac{1}{2}+x_5\right), \\
        C_1&=\eta_{\mathrm{tls}} \frac{1}{8}\left[t_1\left(\frac{1}{2}-x_1\right)-t_2\left(\frac{1}{2}+x_2\right)\right], &C_1^{\prime}&=\eta_{\rm tls} \frac{1}{16}(t_1 - t_2), \\
        C_2&=\eta_{\mathrm{tls}} \frac{1}{8}t_4\left(\frac{1}{2}-x_4\right), &C_2^{\prime}&=\eta_{\rm tls} \frac{1}{16} t_4, \\
        C_3&=-\eta_{\mathrm{tls}} \frac{1}{8}t_5\left(\frac{1}{2}+x_5\right), &C_3^{\prime}&=-\eta_{\rm tls} \frac{1}{16} t_5. \\
        \end{aligned}
        \end{equation}
Note that the tensor SO terms related to $C_i$ and $C_i^{\prime}$ are  omitted in many Skyrme EDFs. Therefore, we use the parameter $\eta_{\rm tls}$ as a switch factor~\cite{Klupfel:2008af} where $\eta_{\rm tls}=1$ and $0$ represent the full inclusion and absence of the $\bm{J}_{ q}^2$ terms, respectively. The parameter $\eta_{\rm tls}$ vanishes ($\eta_{\rm tls}=0$) in the $\rm eS250$ and eS53 EDFs, while $\eta_{\rm tls}=1$ for $\rm eS500_{T}$. For zero-range two-body SO interaction used in the conventional (extended) Skyrme interaction, one has $b_{\rm IS} = 3b_{\rm{IV}} = 3W_0/2$. In this work, we start from the energy density functional and consider $b_{\rm IS}$ and $b_{\rm IV}$ as free and independent parameters~\cite{Reinhard:1995}.

Based on the above energy density functional, nuclear ground-state properties are calculated by solving the HF(B) equations. The single nucleon Hamiltonian entering into the HF(B) equation can be expressed as
\begin{eqnarray}\label{Eq:hq}
\hat{h}_{ q} =-{\bm \nabla} \cdot \frac{\hbar^2}{2m_{ q}^*}{\bm \nabla} +U_{ q}+i\bm{W}_{ q}\cdot(\bm{\sigma}\times \bm \nabla),~  q=\rm n,~p,
\end{eqnarray}
where the single-nucleon fields derived from the energy density $\mathcal{E}$ by
\begin{equation}
\begin{aligned}
\frac{\hbar^2}{2 m_{ q}^*} =\frac{\partial \mathcal{E}}{\partial \tau_{ q}}, ~~U_{ q}=\frac{\partial \mathcal{E}}{\partial \rho_{ q}}-\nabla \cdot \frac{\partial \mathcal{E}}{\partial\left[\nabla \rho_{ q}\right]},~~
\boldsymbol{W}_{ q} =\frac{\partial \mathcal{E}}{\partial \boldsymbol{J}_{ q}} .
\end{aligned}
\end{equation}
From Eq.~(\ref{Eq:Sky-HF}), one can obtain the nucleon effective mass $m_{ q}^*$, i.e.,
\begin{eqnarray}
\frac{\hbar^2}{2m_{ q}^*} = \frac{\hbar^2}{2m} +(B_1+B_4\rho^{\beta}+B_5\rho^{\gamma})\rho -t_q(B_1^{\prime}+B_4^{\prime}\rho^{\beta}+B_5^{\prime}\rho^{\gamma}) \tilde{\rho},
\end{eqnarray}
the single nucleon potential (without Coulomb potential)
\begin{eqnarray}
U_{ q} &=& B_0\rho -t_{ q} B_0^{\prime}\tilde{\rho}+B_1\tau -t_{ q}B_1^{\prime}\tilde{\tau}+\frac{\alpha+2}{2}B_3 \rho^{\alpha+1} -\frac{B_3^{\prime}}{2}(\alpha \tilde{\rho} +2t_{ q} \rho)\rho^{\alpha-1}\tilde{\rho} \notag \\
&&+(\beta+1)B_4\rho^{\beta}\tau +(\gamma+1)B_5\rho^{\gamma}\tau
-(B_4^{\prime}\beta \rho^{\beta}+B_5^{\prime}\gamma \rho^{\gamma})\tilde{\rho}\tilde{\tau}-t_{ q}(B_1^{\prime}+B_4^{\prime}\rho^{\beta}+B_5^{\prime}\rho^{\gamma})\tilde{\tau} \notag \\
 && -\frac{\beta(2\beta+3)B_4 \rho^{\beta-1}-B_5\gamma \rho^{\gamma-1}}{4}(\nabla \rho)^2  -\frac{2B_2+(2\beta+3)B_4\rho^{\beta} -B_5\rho^{\gamma}}{2}\nabla^2 \rho \notag \\
 && -\frac{3\beta B_4^{\prime}\rho^{\beta-1}-\gamma B_5^{\prime}\rho^{\gamma-1}}{4} \nabla \tilde{\rho} \cdot(2t_{ q}\nabla\rho +\nabla \tilde{\rho})-\frac{2B_2^{\prime}+3B_4^{\prime}\rho^{\beta}-B_5^{\prime}\rho^{\gamma}}{2}t_{ q}\nabla^2 \tilde{\rho}\notag \\
 && + \frac{\beta B_4^{\prime}}{2}\rho^{\beta-1}(\nabla \tilde{\rho})^2
+ \frac{\beta B_4^{\prime}}{2}\rho^{\beta-1}\tilde{\rho}\nabla^2 \tilde{\rho}+ \frac{\beta(\beta-1)B_4^{\prime}}{2}\rho^{\beta-2}\tilde{\rho}(\nabla\rho)^2t_{ q}+
\frac{\beta B_4^{\prime}}{2}\rho^{\beta-1}\tilde{\rho}(\nabla^2\rho)t_{ q}  \notag \\
 & &+ \frac{\beta C_2\rho^{\beta-1}+\gamma C_3\rho^{\gamma-1}}{2} \bm{J}^2 + \frac{\beta C_2^{\prime}\rho^{\beta-1}+\gamma C_3^{\prime}\rho^{\gamma-1}}{2} \bm{\tilde{J}}^2 -\frac{b_{\rm{IS}}}{2}\nabla \cdot \bm{J}-t_{ q}\frac{b_{\rm{IV}}}{2}\nabla \cdot \bm{\tilde{J}},
%& &+\frac{\beta^2 B_4^{\prime}}{2}t_q(\nabla \rho)^2 +\frac{\beta B_4^{\prime}}{2}\rho^{\beta} t_q\nabla^2 \rho \\
\label{Eq:CPot}
\end{eqnarray}
and the SO potential
\begin{eqnarray}
\bm{W}_{ q}= \frac{b_{\rm{IS}}}{2}\nabla
\rho+t_{ q} \frac{b_{\rm{IV}}}{2}\nabla(\rho_{\rm n}-\rho_{\rm p})
+ \frac{\alpha_{\rm{J}}+\beta_{\rm{J}}}{2}\bm{J} + t_{ q} \frac{\alpha_{\rm{J}}-\beta_{\rm{J}}}{2}(\bm{J}_{\rm n}-\bm{J}_{\rm p}),
\label{Eq:SOPot}
\end{eqnarray}
where $t_{ q} = 1$ and $-1$ for $ q= n$ and $\rm p$, respectively. The parameters  $\alpha_{\rm J} $ and $\beta_{\rm J}$ in the SO potentials are defined by
\begin{eqnarray}
\alpha_{\rm J} = \alpha_{\rm C}+\alpha_{\rm T}, ~\beta_{\rm J} = \beta_{\rm C} +\beta_{\rm T},
\end{eqnarray}
with the central-exchange parameters given by
\begin{eqnarray}
\alpha_{\rm{C}} = 2C_1 + 2 C_2 \rho^{\beta}  +2 C_3\rho^{\gamma} , ~~\beta_{\rm{C}} = 2C_1^{\prime} + 2 C_2^{\prime} \rho^{\beta}  +2 C_3^{\prime}\rho^{\gamma}.
\end{eqnarray}

\section{Charge and weak form factors}\label{sec:smsubsec2}

The normalized nuclear form factors $F_{\rm C}( q)$ and $F_{\rm W}( q)$ are calculated by folding the nucleon form factor $F_{ t}(q)~( t=n,p)$ and the SO current form factor ($F_{t}^{\rm ls}$) with the intrinsic nucleon electromagnetic form factor $G_{\rm{E/M},t}$ and weak form factor $G^{(\rm W)}_{\rm{E/M},t}$ by~\cite{Reinhard:2021utv}
\begin{eqnarray}
\begin{aligned}
F_{\rm C}(q) &=&\frac{1}{Z} \sum_{t=\rm{p, n}}\left[G_{{\rm E}, t}(q) F_{t}({q})+G_{{\rm M}, t}({ q}) F_{t}^{({\rm l s})}( q)\right],\\
F_{\rm W}({ q}) &=&\sum_{ t=\mathrm{p, n}}\frac{\left[G_{{\rm E}, t}^{(\rm W)}(q) F_{t}({ q})+G_{{\rm M}, t}^{(\rm W)}({ q}) F_{t}^{({\rm l s})}(q)\right]}{Z Q_{\rm p}^{({\rm W})}+N Q_{\rm n}^{({\rm W})}},
\end{aligned}
\end{eqnarray}
where $N(Z)$ is the neutron (proton) number, and  $Q_{\rm p}^{({\rm W})} = 0.0713$ and $Q_{\rm n}^{({\rm W})}=-0.9888$ are proton and neutron weak charges, respectively.
The $G_{{\rm E/M},t}$ are derived from the isospin-coupled Sachs form factors, the relativistic Darwin correction has been included and the center-of-mass corrections are taken into account by simply renormalizing the nucleon mass $m_{\rm N}$ to $(1-1/A)m_{\rm N}$ in the HF calculation (see, e.g., Ref.~\cite{Bender:2003jk} for details). The weak intrinsic nucleon form factors $G^{({\rm W})}_{\rm E/M,t}$ are then determined by~\cite{Reinhard:2021utv}
\begin{eqnarray}
\begin{aligned}
&G_{\rm E/M, n}^{({\rm W})}=Q_{\rm n}^{({\rm W})} G_{\rm E/M, p}+Q_{\rm p}^{({\rm W})} G_{\rm E/M, n}+Q_{\rm n}^{({\rm W})} G_{\rm E/M, s}, \\
&G_{\rm E/M, p}^{({\rm W})}=Q_{\rm p}^{({\rm W})} G_{\rm E/M, p}+Q_{\rm n}^{({\rm W})} G_{\rm E/M, n}+Q_{\rm n}^{({\rm W})} G_{\rm E/M, s},
\end{aligned}
\end{eqnarray}
with the strange-quark electromagnetic form factors $G_{\rm E/M,s}$ given by
\begin{eqnarray}
\begin{aligned}
G_{\rm E, s}({ q}) &=\rho_{\rm s} \frac{\hbar^{2} q^{2} /\left(4 c^{2} m_{\rm N}^{2}\right)}{1+4.97 \hbar^{2} q^{2} /\left(4 c^{2} m_{\rm N}^{2}\right)}, \\
G_{\rm M, s}(q) &=\frac{\kappa_{\rm s}}{(1+\frac{q^2}{12\kappa_{\rm s}}r^2_{\rm M,s})^2}.
\end{aligned}
\end{eqnarray}
Here, the strange quark electric coupling $\rho_{\rm s} = -0.24$~\cite{HAPPEX:2006oqy,Liu:2007yi}, the strange magnetic moment $\kappa_{\rm s} =-0.017$~\cite{Alexandrou:2019olr}, and  the strange magnetic radius
$r^2_{\rm M,s} = -0.015~\rm{fm}^2$~\cite{Alexandrou:2019olr}.

\section{Fitting Process}\label{sec:smsubsec3}

The extended Skyrme interaction has fifteen independent parameters: $t_0\sim t_5$, $x_0\sim x_5$, $\alpha, \beta, \gamma$. For simplicity, we set $\beta=\gamma=1$. The remaining 13 parameters can be analytically determined by 13 macroscopic characteristic quantities of nuclear matter~\cite{Zhang:2015vaa}: the saturation density $\rho_0$, the binding energy per nucleon of symmetric nuclear matter~(SNM) $E_0(\rho_0)$, the incompressibility $K_0$, the magnitude $E_{\rm sym}(\rho_0)$ and density slope $L$ of $E_{\rm sym}(\rho)$ at $\rho_0$, the isoscalar effective mass $m_{\rm s,0}^*$ and isovector effective mass $m_{\rm v,0}^*$ at $\rho_0$, the gradient coefficient $G_{\rm S}$, the symmetry-gradient coefficient $G_{\rm V}$~\cite{Chen2010,Chen:2011ps,Kortelainen:2010hv}, the cross gradient coefficient $G_{\rm SV}$, the Landau parameter $G_0^{\prime}$, the skewness coefficient $J_0$ of SNM and the density curvature parameter $K_{\mathrm{sym}}$ of the $E_{\rm sym}(\rho)$. This mapping ensures that empirical knowledge of nuclear matter can directly inform the Skyrme EDF parameters.

In total, our extended Skyrme EDF framework has 18 adjustable parameters: $\rho_0$, $E_0(\rho_0)$, $K_0$, $J_0$, $E_{\rm sym}(\rho_0)$, $L$, $K_{\rm{sym}}$, $G_0^{\prime}$, $m_{\rm s,0}^*$, $m_{\rm v,0}^*$, $G_{\rm S}$, $G_{\rm V}$, $G_{\rm SV}$, $b_{\rm{IS}}$, $b_{\rm IV}$, $\alpha_{\rm{T}}$, $\beta_{\rm T}$, and the pairing strength $V_0$.  In constructing the eS250 EDF, we set $b_{\rm{IV}}=250~\rm{MeV}\cdot\rm{fm}^5$ and $\alpha_{\rm{T}}=\beta_{\rm{T}}=0$. For the eS500$_{\rm{T}}$, we increase the IVSO strength to $b_{\rm{IV}}=500~\rm{MeV}\cdot\rm{fm}^5$ and treat $\alpha_{\rm{T}}$ and $\beta_{\rm{T}}$ as free parameters. All adjustable parameters except for $V_0$ in eS250 and eS500$_{\rm{T}}$ are determined by
using a two-step strategy: 1) first sample from parameter space based on nuclear structure observables and constraints on nuclear matter; 2) then select parameter sets from these samples according to strict criteria. The pairing parameter $V_0$ is subsequently adjusted to be $V_0 = -230\,\mathrm{MeV}\cdot\mathrm{fm}^3$ to reproduce the experimental average pairing gap in $^{120}{\rm Sn}$~\cite{Bender:2000xk}, using HFB calculations~\cite{NavarroPerez:2017bxq} based on the eS250.

In the sampling process, a Markov Chain Monte Carlo (MCMC) approach is implemented to sample the parameters according to a likelihood function
%with commonly used Gaussian form
\begin{eqnarray}
\label{Eq:likelihood}
\mathcal{L}(\bm{p}) \propto \exp \left\{-\sum_{i} \frac{\left[\mathcal{O}_{i}(\boldsymbol{p})-\mathcal{O}_{i}^{\exp }\right]^{2}}{2 \sigma_{i}^{2}}\right\},
\end{eqnarray}
where $\mathcal{O}_{i}(\boldsymbol{p})$ is the model prediction for the $i$-th observable based on a given parameter set $\boldsymbol{p}$,  $\mathcal{O}_{i}^{\exp}$ is the corresponding experimental data, and $\sigma_{i}$ represents the adopted error weighting the importance of the $i$-th observable. The experimental data (see Tab.~\ref{Tab:Data}) considered in the sampling procedure include (i) the binding energies $E_{\rm B}$ of $^{16}{\rm O}$, $^{40,48}{\rm Ca}$, $^{56,68}{\rm Ni}$, $^{100,132}{\rm Sn}$, and $^{208}{\rm Pb}$~\cite{Wang:2021xhn,Klupfel:2008af}; (ii) the charge rms radii $r_{\rm c}$ of $^{16}{\rm O}$, $^{40,48}{\rm Ca}$, $^{56}{\rm Ni}$, and $^{208}{\rm Pb}$~\cite{Angeli:2013epw,Klupfel:2008af,LeBlanc:2005ik}; (iii) the diffraction radii and surface thicknesses of
$^{16}{\rm O}$, $^{40,48}{\rm Ca}$, $^{208}{\rm Pb}$~\cite{Klupfel:2008af}; (iv) the excitation energy of giant monopole resonance in $^{208}{\rm Pb}$~\cite{Patel:2013uyt,Howard:2020pnz,Youngblood:1999zza,Zhang:2022bni}; (iv) SO splittings for the $1p$ neutrons and proton in $^{16}{\rm O}$, $1f$ neutrons in
$^{40,48}{\rm Ca}$, $1f$ neutrons in $^{56}{\rm Ni}$, $2d$ neutrons in $^{132}{\rm Sn}$, and $2f$ protons,  $3p$ neutrons and $2f$ neutrons in $^{208}{\rm Pb}$~\cite{Isakov:2002jv,Schwierz:2007ve}; (v) Charge-Weak form factor difference $\Delta F_{\rm CW}$ in
$^{48}{\rm Ca}$ and $^{208}{\rm Pb}$ measured by CREX and PREX collaborations, respectively~\cite{CREX:2022kgg,PREX:2021umo}.
The experimental values together with the adopted errors used in the sampling procedure are listed in Tab.~\ref{Tab:Data}.
Besides, we include the constraints on pressure of symmmetric nuclear matter  in the density region $2\rho_0 <\rho<4.6\rho_0$ obtained from analyzing flow data in heavy-ion collisions~\cite{Danielewicz:2002pu}
by taking $\mathcal{O}_i^{\exp}$ to be the central values and $\sigma_i$ to be half widths of the constraint band. In the construction of eS250, the predicted uncertainty band for neutron matter equation of state by ab initio calculations reported in Ref.~\cite{Huth:2020ozf} is also considered.

In the selection procedure, the criteria for eS250 include  (i) reproduce the binding energies of
$^{40,48}{\rm Ca}$, $^{88}{\rm Sr}$, $^{56,68}{\rm Ni}$, $^{90}{\rm Zr}$, $^{100,132}{\rm Sn}$ and $^{208}{\rm Pb}$ within $1\%$ relative deviations; (ii) reproduce the charge radii of
$^{40,48}{\rm Ca}$, $^{88}{\rm Sr}$, $^{56}{\rm Ni}$, $^{90}{\rm Zr}$, and $^{208}{\rm Pb}$ within $1\%$ relative deviations; (iii) ensure
the relative deviations from experimental SO splittings in $^{16}{\rm O}$, $^{40,48}{\rm Ca}$, $^{56}{\rm Ni}$, $^{132}{\rm Sn}$ and $^{208}{\rm Pb}$ are less than 50\%; (iv) ensures the predicted pressure in symmetric nuclear matter falls within the constraint band from flow data in HICs; (v) the predicted electric dipole polarizabilities $\alpha_{\rm D}$ and $\Delta F_{\rm CW}$ should be consistent with experimental data within $1\sigma$ error and (vi) the neutron matter equation of state
should be consistent with the predictions of ab initio calculations~\cite{Huth:2020ozf}. For the selection of $\mathrm{eS500_{\rm T}}$, the constraints from both the electric dipole polarizability and the pure neutron matter equation of state are removed compared to eS250.

After the construction of the eS250 EDF, to more clearly demonstrate the effects of the IVSO interaction, we change the $b_{\rm{IV}}$ value to be $ 53~\rm{MeV}\cdot\rm{fm}^5\approx b_{\rm{IS}}/3$ while keep all other parameters unchanged. The resulting EDF is named ``eS53''. The detailed  values of parameters and the coefficients $B_i$ and $C_i$ [see Eq. (~\ref{Eq:Sky-HF})] for the three newly constructed extended Skyrme EDFs, eS53, eS250 and eS500$_{\rm{T}}$ are listed in Tab.~\ref{Tab:BCpara}.

\bibliography{Refsupp}

\begin{table*}[htbp]
\centering
\caption{Experimental data and adopted errors used in the fitting. The table lists nuclear structure observables including binding energy $E_{\rm{B}}$~\cite{Wang:2021xhn,Klupfel:2008af}, charge radius $r_{\rm c}$~\cite{Angeli:2013epw,Klupfel:2008af,LeBlanc:2005ik}, the diffraction radius
$R_{\rm d}$~\cite{Klupfel:2008af}, surface thickness $\sigma$~\cite{Klupfel:2008af}, SO splitting $\Delta \epsilon_{\rm ls}$~\cite{Isakov:2002jv,Schwierz:2007ve},  excitation energy of giant monopole resonance $E_{\rm{GMR}}$ in $^{208}$Pb~\cite{Zhang:2022bni,Patel:2013uyt,Youngblood:1999zza,Howard:2020pnz}, and charge-weak form factor difference $\Delta F_{\rm{CW}}$ in $^{48}$Ca and $^{208}$Pb measured by CREX and PREX collaborations~\cite{CREX:2022kgg,PREX:2021umo}, respectively.
The second line shows the globally adopted error for each observable. That error is multiplied for each observable by a further integer weight factor given in the parenthesis next to the data value.}
\label{Tab:Data}
\renewcommand{\arraystretch}{1.1}
\begin{tabular*}{\textwidth}{@{\extracolsep{\fill}}l*{7}{c}@{}}
\hline\hline
\rm{Nuclei}
& \multicolumn{1}{c}{$E_{\rm{B}}$}
& \multicolumn{1}{c}{$r_{\rm c}$}
& \multicolumn{1}{c}{$R_{{\rm d}}$}
& \multicolumn{1}{c}{$\sigma$}
& \multicolumn{1}{c}{$\Delta\epsilon_{\rm ls}$}
& \multicolumn{1}{c}{$E_{\rm GMR}$}
& \multicolumn{1}{c}{$\Delta F_{\rm CW}$} \\
& \multicolumn{1}{c}{(1 MeV)}
& \multicolumn{1}{c}{(0.02 fm)}
& \multicolumn{1}{c}{(0.04 fm)}
& \multicolumn{1}{c}{(0.04 fm)}
& \multicolumn{1}{c}{(10\%)}
& \multicolumn{1}{c}{(0.074 MeV)}
& \multicolumn{1}{c}{(0.002)} \\
\hline
${}^{16}$O      & $-127.620 (1)$  & $2.701(1)$ & $2.777(1)$ & $0.839(1)$  & $6.32(3)$ &  &             \\
                &                 &            &            &             & $6.17(3)$ &            &             \\
${}^{40}$Ca     & $-342.051 (1)$  & $3.478(1)$ & $3.845(1)$ & $0.978(1)$  & $6.80(1)$ &  &             \\
${}^{48}$Ca     & $-415.990 (1)$  & $3.477(1)$ & $3.964(1)$ & $0.881(1)$  & $8.80(1)$ &    & $0.0277(1)$ \\
${}^{56}$Ni     & $-483.990 (1)$  & $3.750(1)$ &          &           & $7.16(1)$ &  &             \\
${}^{68}$Ni     & $-590.430 (1)$  &          &          &           &         &          &             \\
${}^{100}$Sn    & $-825.800 (1)$  &          &          &           &         &          &             \\
${}^{132}$Sn    & $-1102.900(1)$  &          &          &           & $1.66(3)$ &  &             \\
${}^{208}$Pb    & $-1636.446(1)$  & $5.501(1)$ & $6.776(1)$ & $0.913(1)$  & $1.46(3)$ & $13.614(1)$ & $0.0410(1)$ \\
                &                 &            &            &             & $0.89(3)$ &            &             \\
                &                 &            &            &             & $2.13(3)$ &            &             \\
\hline\hline
\end{tabular*}
\end{table*}

\begin{table*}[t]  % 使用table*环境实现跨双栏
\centering
\caption{Model parameters and coefficients for the eS250, eS53, and eS500$_{\rm T}$ energy density functionals.  }
\label{Tab:BCpara}
\renewcommand{\arraystretch}{1.2} % 增加行高提升可读性
\begin{tabular*}{\textwidth}{@{\extracolsep{\fill}}cccc|cccc@{}}
\hline\hline
quantity & eS250 & $\rm eS53$ & $\rm eS500_T$ & quantity & eS250 & $\rm eS53$ & $\rm eS500_T$ \\
\hline
$t_0~(\mathrm{MeV}\cdot \mathrm{fm}^{5})$  & -1803.9 & -1803.9 & -1594.0  & $b_{\rm{IS}}~(\mathrm{MeV}\cdot \mathrm{fm}^{5})$ & 160.461 & 160.461 & 134.149 \\
$t_1~(\mathrm{MeV}\cdot \mathrm{fm}^{5})$ & 450.79 & 450.79 & 589.01  & $b_{\rm{IV}}~(\mathrm{MeV}\cdot \mathrm{fm}^{5})$ & 250 & 53 & 500  \\
$t_2~(\mathrm{MeV}\cdot \mathrm{fm}^{5})$ & 291.3 & 291.3 & -754.1   & $\rho_0~(\rm{fm}^{-3})$ & 0.15881 & 0.15881 & 0.150853 \\
$t_3~(\mathrm{MeV}\cdot \mathrm{fm}^{3+3\alpha})$ & 12514 & 12514 & 12135  & $E_0~(\rm{MeV})$ & -16.1209 & -16.1209 & -15.8993 \\
$t_4~(\mathrm{MeV}\cdot \mathrm{fm}^{3+3\beta})$ & -922.51 & -922.51 & -1864.1  & $K_0~(\rm{MeV})$ & 238.125 & 238.125 & 237.424  \\
$t_5~(\mathrm{MeV}\cdot \mathrm{fm}^{3+3\gamma})$ & -1090.7 & -1090.7 & 5491.5  & $J_0~(\rm{MeV})$ & -393.323 & -393.323 & -467.697  \\
$x_0$ & 0.26206 & 0.26206 & -0.31013  & $E_{\rm{sym}}(\rho_0)~(\rm{MeV})$ & 33.9279 & 33.9279 & 35.6061  \\
$x_1$ & -0.95171 & -0.95171 & 0.07722  & $L~(\rm{MeV})$ & 58.4409 & 58.4409 & 84.7663  \\
$x_2$ & -2.249 & -2.249 & -0.83788  & $K_{\mathrm{sym}}~(\rm{MeV})$ & -111.833 & -111.833 & -201.929  \\
$x_3$ & 0.07903 & 0.07903 & -1.0964  & $m_{\rm s,0}/m$ & 0.92297 & 0.92297 & 0.85291  \\
$x_4$ & -1.5503 & -1.5503 & 0.36685  & $m_{\rm v,0}/m$ & 0.77078 & 0.77078 & 0.69718  \\
$x_5$ & -1.8642 & -1.8642 & -1.0391  & $G_0^{\prime}$ & 0.75906 & 0.75906 & 0.71105  \\
$\alpha$ & 0.33162 & 0.33162 & 0.44449  & $G_\mathrm{S}~(\mathrm{MeV}\cdot \mathrm{fm}^{5})$ & 81.1891 & 81.1891 & 50.8546  \\
$\beta$ & 1 & 1 & 1  & $G_{\mathrm{V}}~(\mathrm{MeV}\cdot \mathrm{fm}^{5})$ & -26.4039 & -26.4039 & 6.05963  \\
$\gamma$ & 1 & 1 & 1  & $G_{\rm SV}~(\mathrm{MeV}\cdot \mathrm{fm}^{5})$ & 19.2338 & 19.2338 & -30.4694  \\
\hline
$B_0~(\mathrm{MeV}\cdot \mathrm{fm}^{3})$ & -1352.90 & -1352.90 & -1195.54 & $B_0^{\prime}~(\mathrm{MeV}\cdot \mathrm{fm}^{3})$ & -687.32 & -687.32 & -151.33 \\
$B_1~(\mathrm{MeV}\cdot \mathrm{fm}^{5})$ & 11.768 & 11.768 & 32.740 & $B_1^{\prime}~(\mathrm{MeV}\cdot \mathrm{fm}^{5})$ & 38.234 & 38.234 & 10.647 \\
$B_2~(\mathrm{MeV}\cdot \mathrm{fm}^{5})$ & 163.16 & 163.16 & 204.51 & $B_2^{\prime}~(\mathrm{MeV}\cdot \mathrm{fm}^{5})$ & -70.024 & -70.024 & 79.673 \\
$B_3~(\mathrm{MeV}\cdot \mathrm{fm}^{3+3\alpha})$ & 1564.21 & 1564.21 & 1516.83 & $B_3^{\prime}~(\mathrm{MeV}\cdot \mathrm{fm}^{3+3\alpha})$ & 603.82 & 603.82 & -603.13 \\
$B_4~(\mathrm{MeV}\cdot \mathrm{fm}^{5+3\beta})$ & -172.97 & -172.97 & -349.51 & $B_4^{\prime}~(\mathrm{MeV}\cdot \mathrm{fm}^{5+3\beta})$ & 121.11 & 121.11 & -201.98 \\
$B_5~(\mathrm{MeV}\cdot \mathrm{fm}^{5+3\beta})$ & 167.48 & 167.48 & 289.60 & $B_5^{\prime}~(\mathrm{MeV}\cdot \mathrm{fm}^{5+3\beta})$ & -185.99 & -185.99 & 370.02 \\
$C_1~(\mathrm{MeV}\cdot \mathrm{fm}^{5})$ & 0 & 0 & -0.723 & $C_1^{\prime}~(\mathrm{MeV}\cdot \mathrm{fm}^{5})$ & 0 & 0 & 83.947 \\
$C_2~(\mathrm{MeV}\cdot \mathrm{fm}^{5+3\beta})$ & 0 & 0 & -31.026 & $C_2^{\prime}~(\mathrm{MeV}\cdot \mathrm{fm}^{5+3\beta})$ & 0 & 0 & -116.50 \\
$C_3~(\mathrm{MeV}\cdot \mathrm{fm}^{5+3\beta})$ & 0 & 0 & 370.02 & $C_3^{\prime}~(\mathrm{MeV}\cdot \mathrm{fm}^{5+3\beta})$ & 0 & 0 & -343.22 \\
\hline\hline
\end{tabular*}
\end{table*}

\begin{table*}[t]
\centering
\caption{Predictions for ground- and collective excited-state properties of $^{48}$Ca and $^{208}$Pb.
Results from eS250, eS53 and eS500$_{\rm{T}}$ results are presented for the charge radii $r_{\rm c}$, electric-weak charge form factor differences $\Delta F_{\rm{CW}}$, neutron skin thicknesses $\Delta r_{\rm{np}}$, and electric dipole polarizabilitie $\alpha_{\rm{D}}$ of $^{48}$Ca and $^{208}$Pb, as well as the breathing mode energy $E_{\rm{GMR}}$ in $^{208}$Pb. Superscripts  ``48'' and ``208'' indicate results for $^{48}$Ca and $^{208}$Pb, respectively.  Experimental data are included for comparison.}
\label{Tab:Parameters}
\renewcommand{\arraystretch}{1.1}
\begin{tabular*}{\textwidth}{@{\extracolsep{\fill}}cccccccccc@{}}
\hline \hline
quantity & eS250 & $\rm eS53$ & $\rm eS500_T$ & exp & quantity & eS250 & $\rm eS53$ & $\rm eS500_T$ & exp \\
\hline
$r_{\rm c}^{48}~(\rm{fm})$ & 3.496 & 3.481 & 3.506 & 3.477~\cite{Angeli:2013epw} & $r_{\rm c}^{208}~(\rm{fm})$ & 5.483 & 5.483 & 5.501 & 5.501~\cite{Angeli:2013epw} \\
$\Delta F_{\rm{CW}}^{48}$ & 0.0331 & 0.0428 & 0.0289 & $0.0277\pm0.0055$~\cite{CREX:2022kgg} & $\Delta F_{\rm{CW}}^{208}$ & 0.0292 & 0.0295 & 0.0421 & $0.041\pm0.013$~\cite{PREX:2021umo} \\
$\Delta r_{\rm{np}}^{48}~(\rm{fm})$ & 0.130& 0.180 & 0.087 & & $\Delta r_{\rm{np}}^{208}~(\rm{fm})$ & 0.195 & 0.198 & 0.279 & \\
$\alpha_{\rm{D}}^{48}~(\rm{fm}^3)$ & 2.25 & 2.31 & 3.03 & $2.07\pm0.22$~\cite{Birkhan:2016qkr} & $\alpha_{\rm{D}}^{208}~(\rm{fm}^3)$ & 20.06 & 19.78 & 24.52 & $19.60\pm0.60$~\cite{Tamii2011a,Roca-Maza:2015eza} \\
$E_{\rm GMR}^{208}~(\rm{MeV})$ & 13.78 & 13.78 & 13.50 & $13.614\pm0.074$~\cite{Zhang:2022bni,Patel:2013uyt,Youngblood:1999zza,Howard:2020pnz} & $E_{\rm GMR}^{48}~(\rm{MeV})$ & 20.39 & 20.29 & 18.51 & $19.88\pm0.16$~\cite{Anders:2013rea}  \\
\hline \hline
\end{tabular*}
\end{table*}

\begin{figure*}[h]
  \centering
  \includegraphics[width=\textwidth]{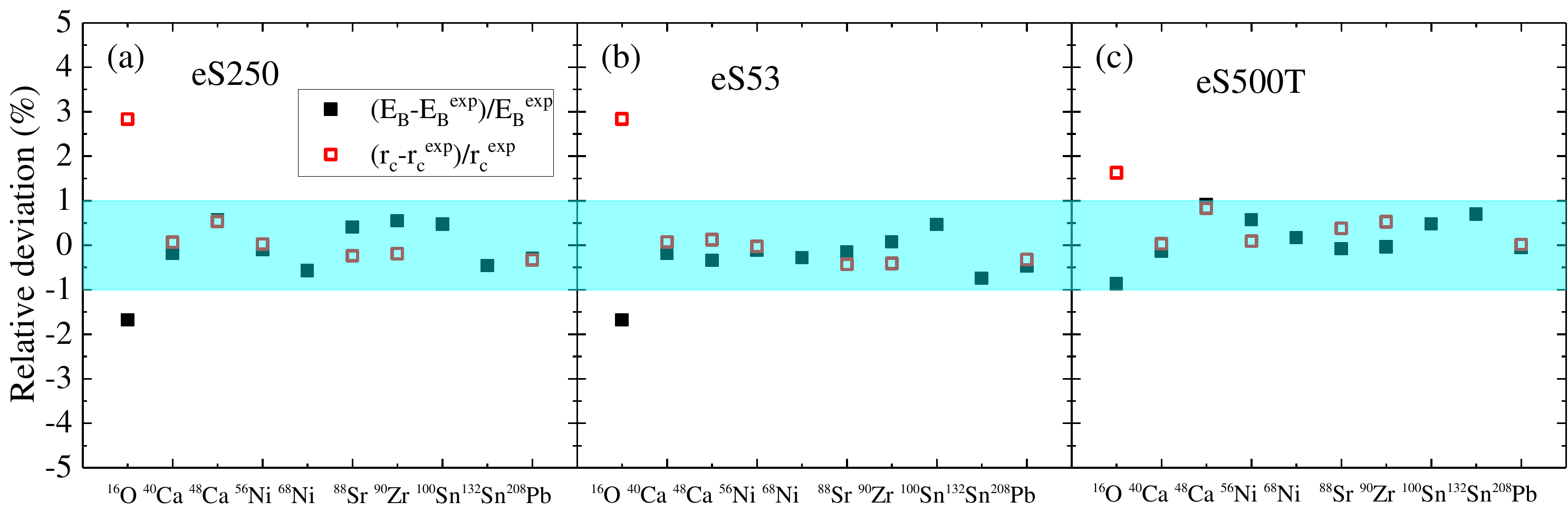}
  \par
  \vspace{1cm}
  \includegraphics[width=0.7\linewidth]{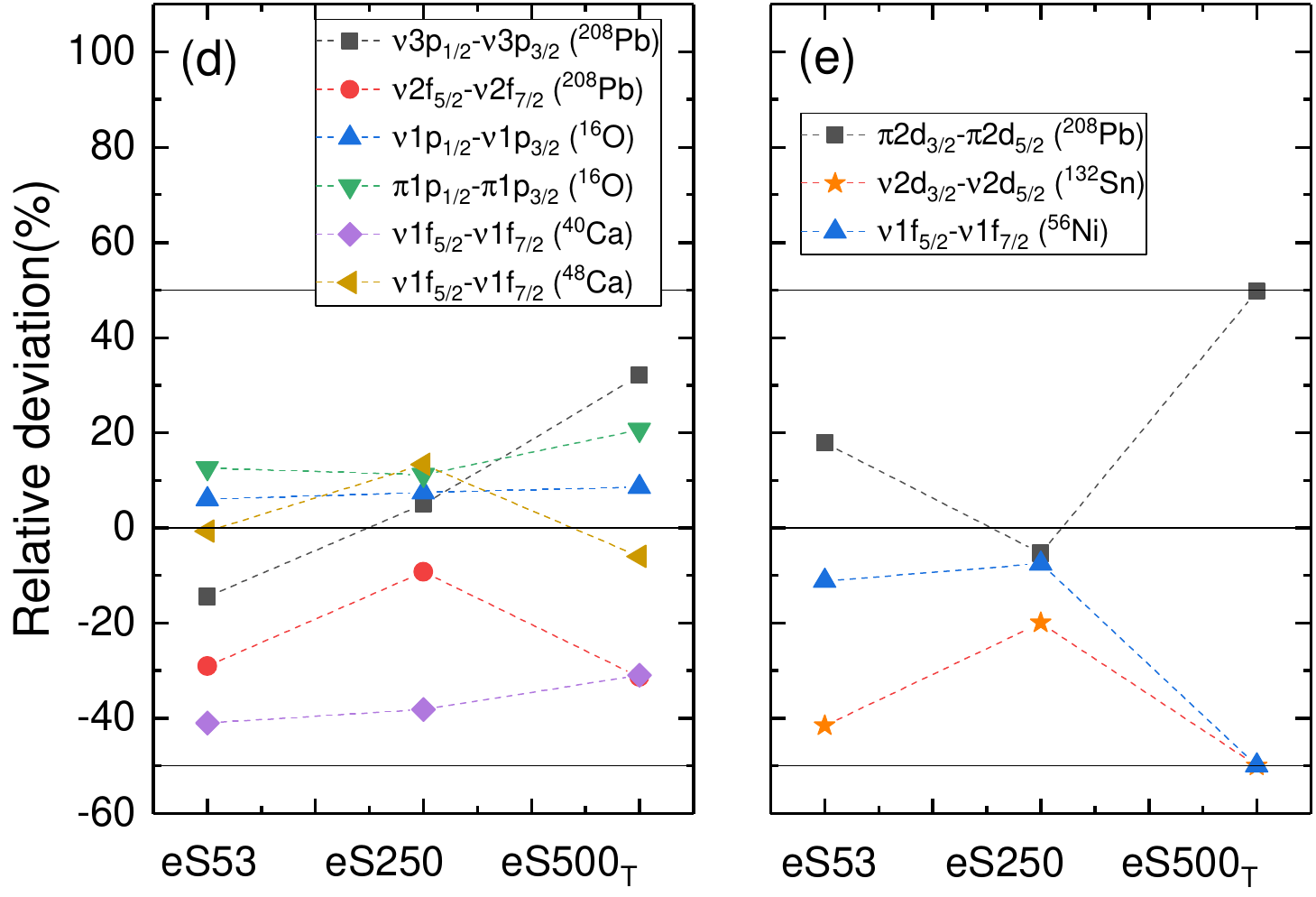}
  \caption{Benchmarking  eS250, eS53, and eS500$_{\rm T}$ against nuclear ground-state observables. (a-c) Relative deviations of binding energies ($E_{\rm B}$) and charge radii ($r_{\rm c}$) for selected closed-shell nuclei ($^{16}$O, $^{40}$Ca, $^{48}$Ca, $^{56}$Ni, $^{68}$Ni, $^{88}$Sr, $^{90}$Zr, $^{100}$Sn, $^{132}$Sn, $^{144}$Sm, and $^{208}$Pb), showing agreement within $\pm1\%$ for most systems except light nuclei like $^{16}$O where mean-field approximations are less valid. The shaded bands indicates $\pm1\%$. (d-e) Relative deviations of SO splittings in doubly magic nuclei ($^{16}$O, $^{40}$Ca, $^{48}$Ca, $^{56}$Ni, $^{132}$Sn and $^{208}$Pb), demonstrating reproduction of experimental trends within $\pm50\%$. Experimental data are from Refs.~\cite{Wang:2021xhn,Angeli:2013epw,Isakov:2002jv, Schwierz:2007ve}. }
  \label{Fig:EbRc-SOsplit}
\end{figure*}

\begin{figure*}[h]
  \centering
  \includegraphics[width=0.8\linewidth]{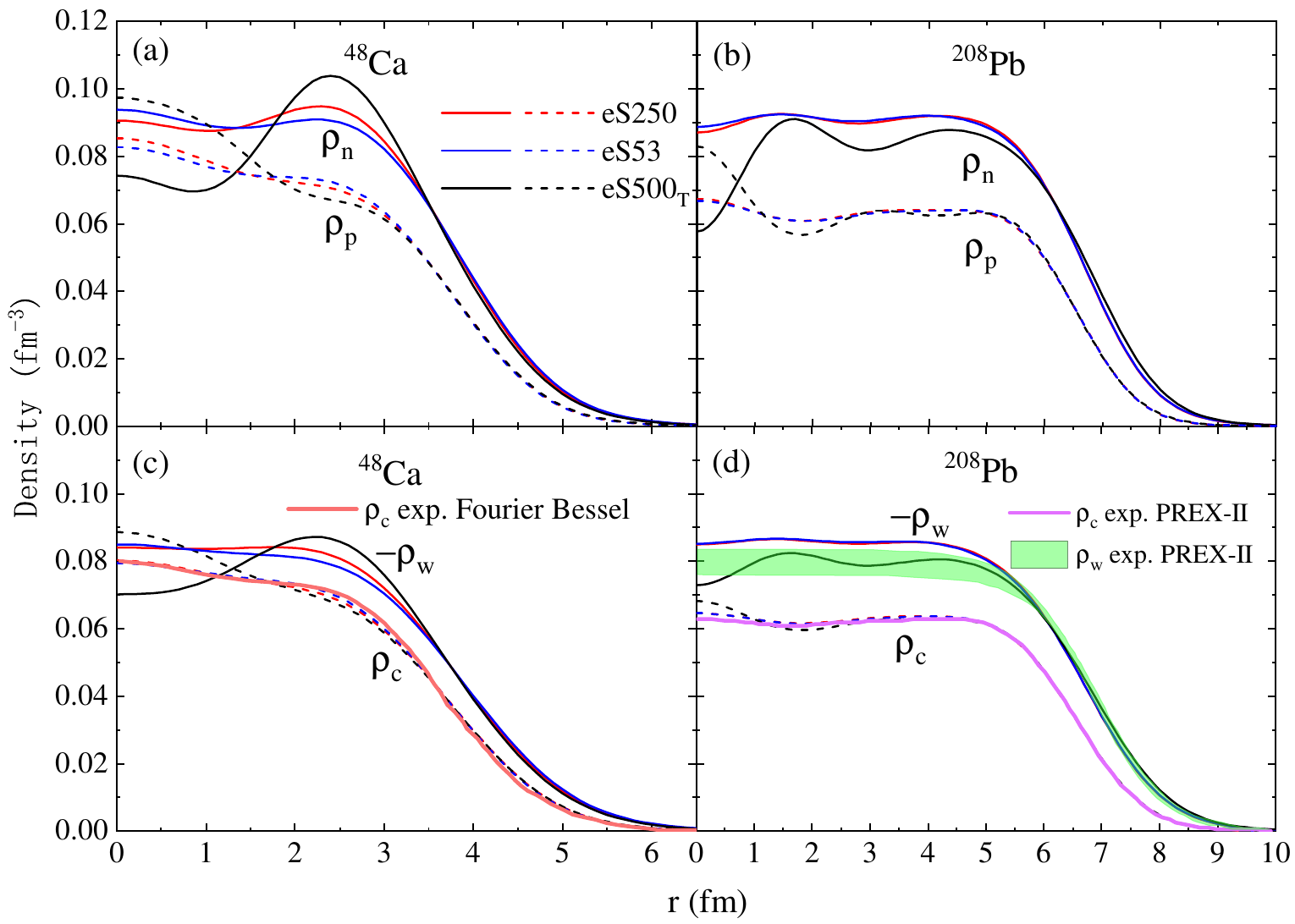}
  \caption{Radial density distributions of $^{48}$Ca (left panels) and $^{208}$Pb (right panels) calculated using three extended Skyrme energy density functionals: eS53, eS250, and eS500T.
The upper panels show neutron ($\rho_{\rm n}$, solid lines) and proton ($\rho_{\rm p}$, dashed lines) densities.
The lower panels display the weak charge density ($-\rho_{\rm w}$, solid lines) and charge density ($\rho_{\rm c}$, dashed lines).
For $^{48}$Ca, the experimental charge density extracted from electron scattering~\cite{DeVries:1987atn} is included in panel (c) for comparison.
For $^{208}$Pb, the experimental charge and weak charge densities inferred from the PREX-2 experiment~\cite{PREX:2021umo} are indicated by shaded bands in panel (d).
}
\label{Fig:rhonpwc}
\end{figure*}

% \begin{figure*}[h]
%   \centering
%   \includegraphics[width=1.0\linewidth]{SFig3Ca48elevel.pdf}
%   \caption{Single-particle spectra in $^{48}{\rm Ca}$. Neutron and proton levels calculated using the eS53, eS250, eS500$_{\rm{T}}$ energy density functionals; deeply bound states (below $-50$ MeV) are omitted.}
% \label{Fig:Ca48elevel}
% \end{figure*}

\begin{figure*}[h]
  \centering
  \includegraphics[width=1.0\linewidth]{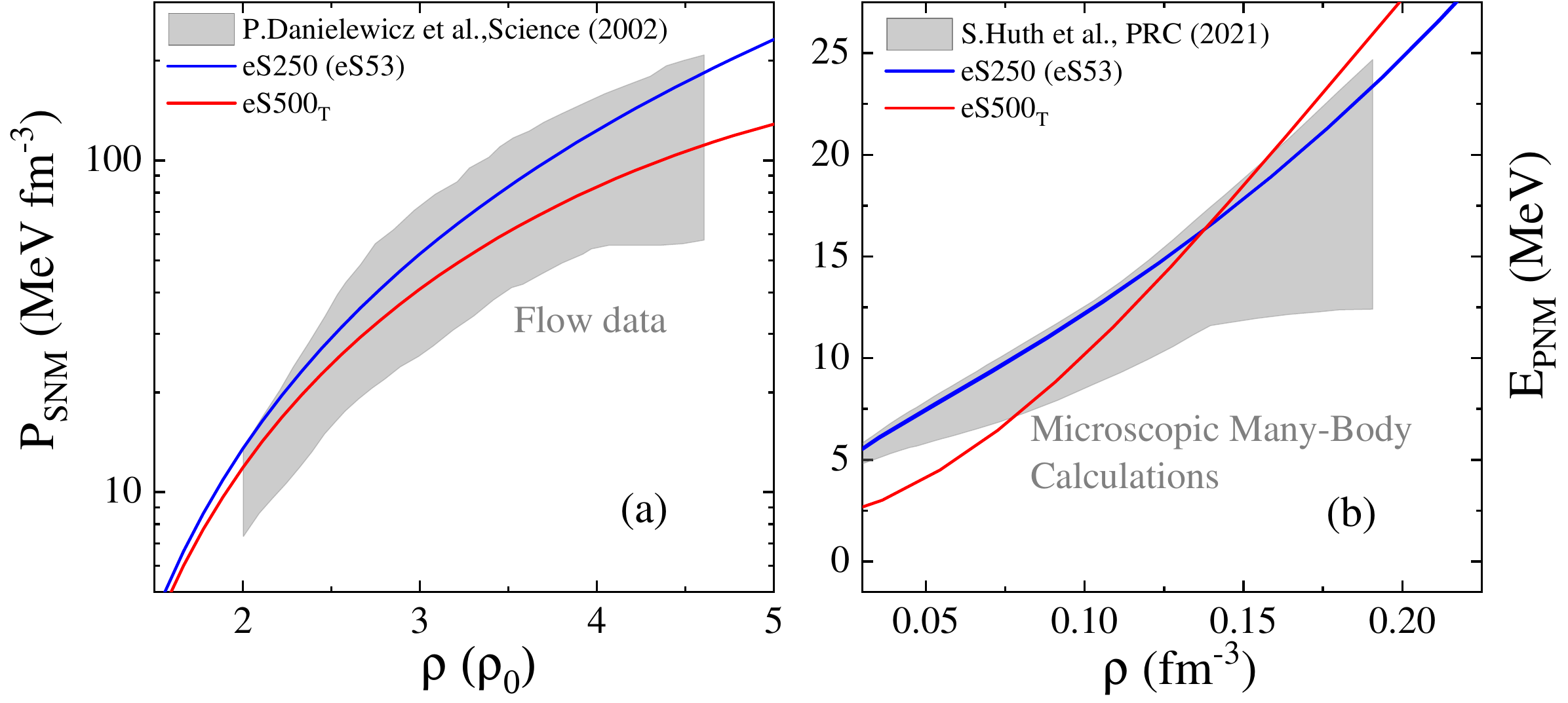}
  \caption{Nuclear equation of state. (a) Pressure $P_{\rm SNM}(\rho)$ in symmetric nuclear matter and (b) energy per neutron $E_{\rm PNM}(\rho)$ in pure neutron matter, from eS250 and eS500$_{\rm T}$. Shaded bands: constraints from heavy-ion collision flow data~\cite{Danielewicz:2002pu} (a) and predictions of microscopic many-body theories~\cite{Huth:2020ozf} (b).}
\label{Fig:EpnmPsnm}
\end{figure*}

\begin{figure*}[h]
  \centering
  \includegraphics[width=0.8\linewidth]{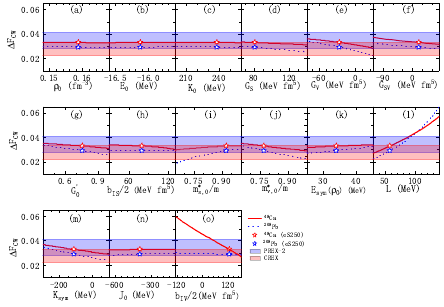}
  \caption{Correlation analysis of model parameters. The $\Delta F_{\rm{CW}}$ in $^{208}{\rm Pb}$ (dashed lines) and $^{48}{\rm Ca}$ (solid lines) from the $\rm eS250$ EDF by varying individually $\rho_0$ (a), $E_0$ (b), $K_0$ (c), $G_{\rm{S}}$ (d), $G_{\rm{V}}$ (e), $b_{\rm{IS}}$ (f), $m_{\rm s,0}^*/m$ (g), $m_{\rm v,0}^*/m$ (h), $E_{\rm sym}(\rho_0)$ (i), $L$ (j), $b_{\rm{IV}}$ (k), $\alpha_{\rm{T}}$ (l) and $\beta_{\rm{T}}$ (m). The PREX-2 and CREX measurements with $1\sigma$ error are shown by the light blue and red bands, respectively. The $\rm eS250$ predictions are indicated by open stars. $\Delta F_{\rm CW}^{208}$ is sensitive primarily to the symmetry-energy slope $L$, whereas $\Delta F_{\rm CW}^{48}$ correlates positively with $L$ and negatively with $b_{\rm IV}$. All other parameters have only minor effects.}
\label{Fig:CorrelationSHF}
\end{figure*}

\begin{figure*}[h]
  \centering
  \includegraphics[width=0.8\linewidth]{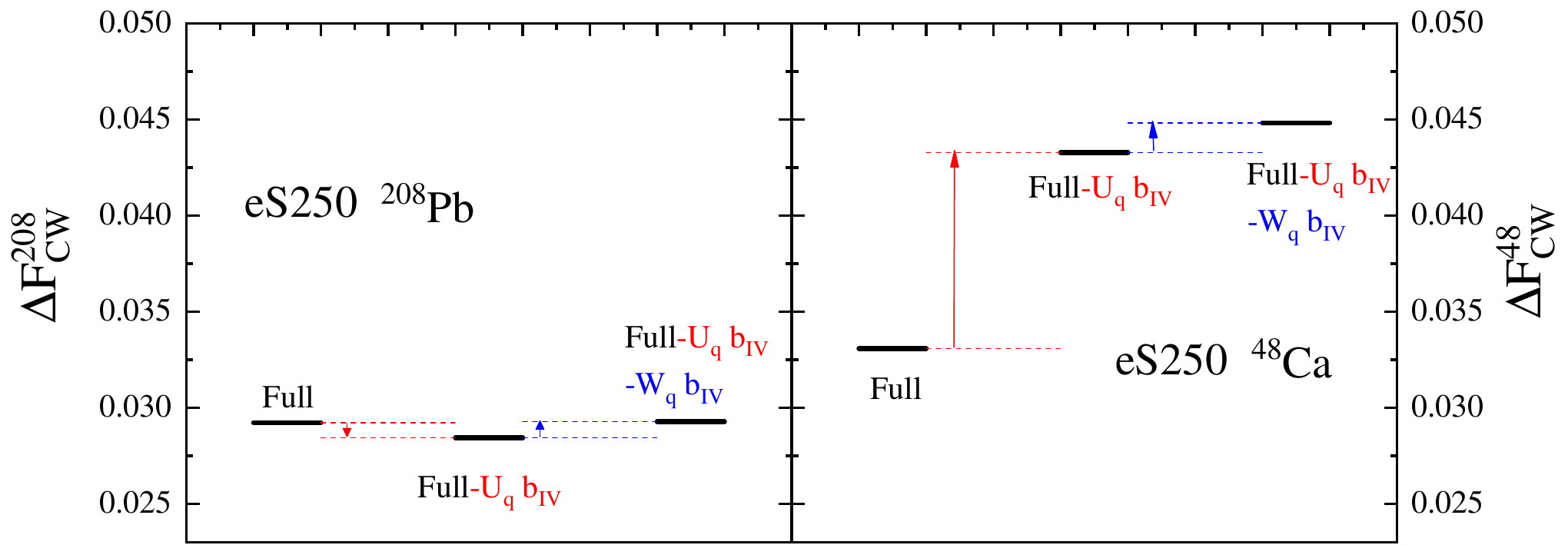}
  \par
  \vspace{1cm}
  \includegraphics[width=0.8\linewidth]{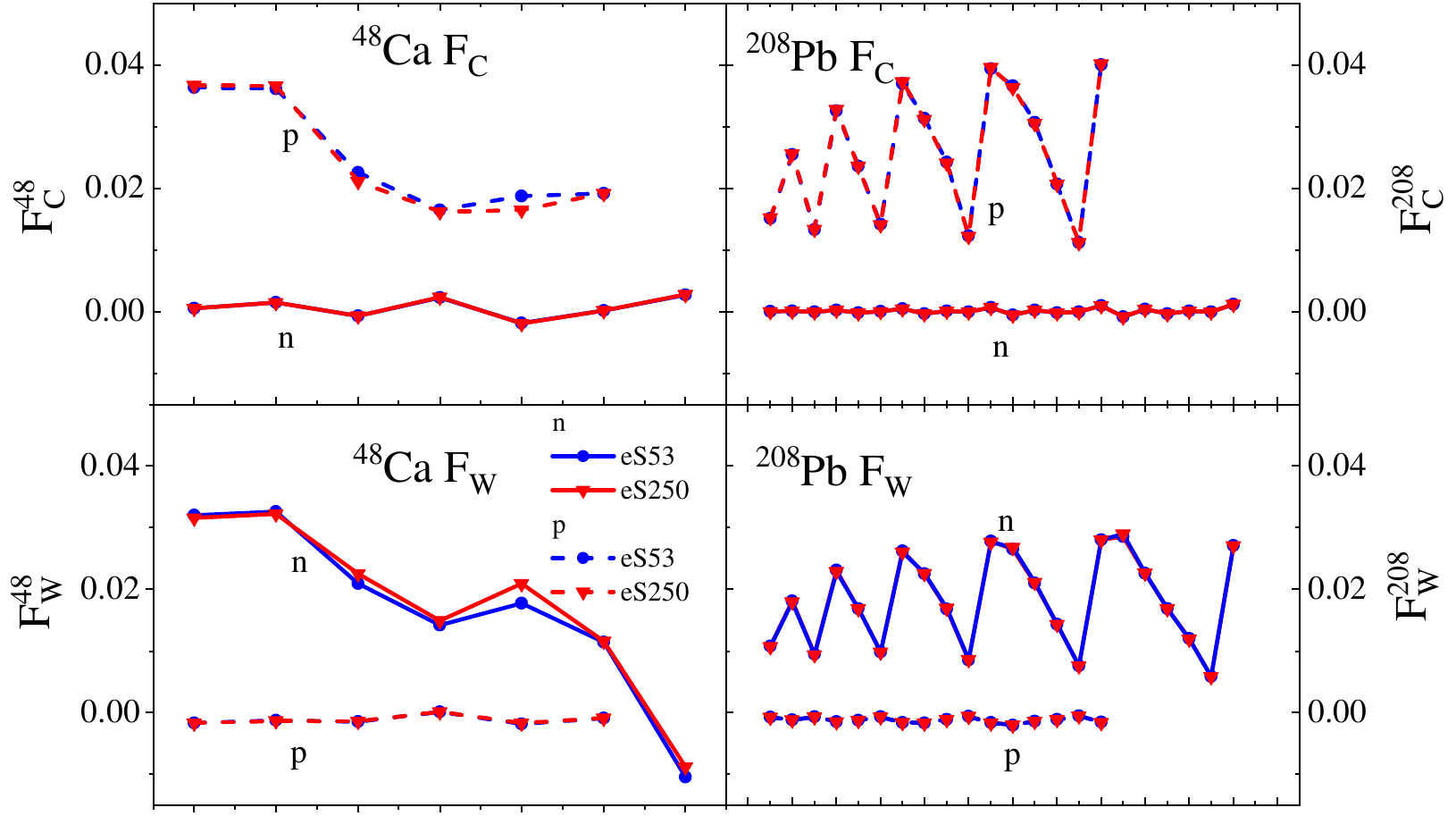}
  \caption{Comparative analysis of the effects of the IVSO interaction on the charge and weak form factors in $^{48}$Ca and $^{208}$Pb. The upper panels show the charge-weak form factor difference $\Delta F_{\rm CW}$ calculated using the eS250 EDF while selectively switching on or off the IVSO contribution in: (1) both central and SO potentials; (2) only the SO potential; and
(3) both potentials (full eS250). The
middle and lower panels display, respectively, the single-orbital contributions to the charge and weak form factor, $F_{\rm C}$ and $F_{\rm W}$, calculated with both the eS53 and eS250 EDFs. For $^{208}$Pb, the analyzed orbitals (from left to right)  include: $\rm 1s_{1/2}$, $\rm 1p_{3/2}$, $\rm 1p_{1/2}$, $\rm 1d_{5/2}$, $\rm 1d_{3/2}$, $\rm 2s_{1/2}$, $\rm 1f_{7/2}$, $\rm 1f_{5/2}$, $\rm 2p_{3/2}$, $\rm 2p_{1/2}$, $\rm 1g_{9/2}$, $\rm 1g_{7/2}$, $\rm 2d_{5/2}$, $\rm 2d_{3/2}$, $\rm 3s_{1/2}$, $\rm 1h_{11/2}$, $\rm 1h_{9/2}$, $\rm 2f_{7/2}$, $\rm 2f_{5/2}$, $\rm 3p_{3/2}$, $\rm 3p_{1/2}$, and $\rm 1i_{13/2}$, while for $^{48}$Ca, the analyzed orbitals (from left to right) are $\rm 1s_{1/2}$, $\rm 1p_{3/2}$, $\rm 1p_{1/2}$, $\rm 1d_{5/2}$, $\rm 1d_{3/2}$, $\rm 2s_{1/2}$, $\rm 1f_{7/2}$.}
\label{Fig:SFig6IVSOUWls}
\end{figure*}

\begin{figure*}[h]
  \centering
  \includegraphics[width=0.8\linewidth]{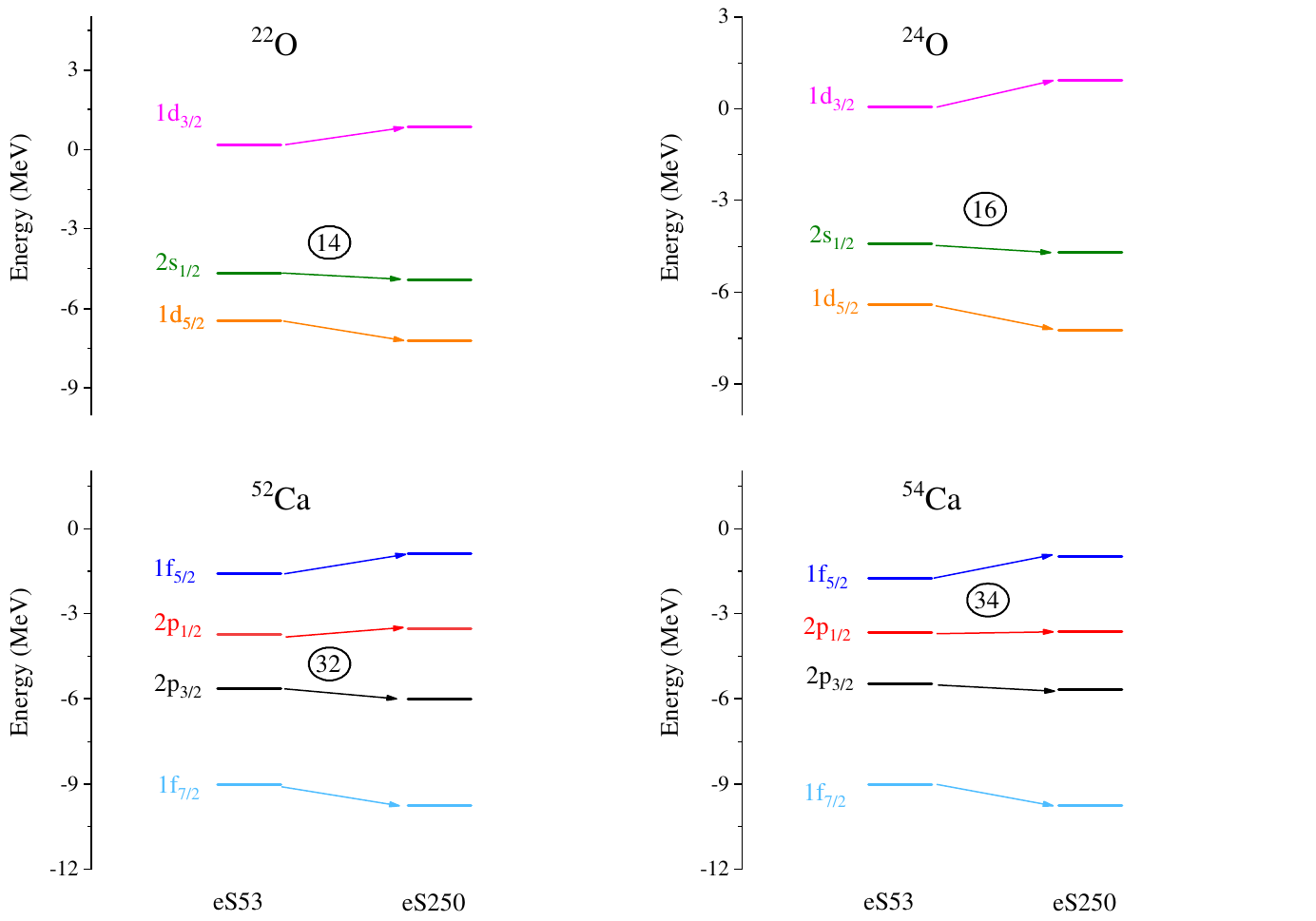}
\caption{Key orbital evolutions leading to new magic numbers $N=14$, $16$, $32$, and $34$ in $^{22}$O, $^{24}$O, $^{52}$Ca, and $^{54}$Ca. Each subpanel compares the last occupied and first unoccupied neutron single-particle levels from eS53 (left) and eS250 (right): (a) in $^{22}$O, an enhanced gap of $\nu1d_{5/2}$ and $\nu2s_{1/2}$ establishes the $N=14$ shell closure; (b) in $^{24}$O, an enhanced $\nu2s_{1/2}$-$\nu1d_{3/2}$ gap establishes the $N=16$ shell closure; (c) in $^{52}$Ca, an increased $\nu2p_{3/2}$-$\nu2p_{1/2}$ gap establishes the $N=32$ shell closure; (d) in $^{54}$Ca, the $\nu1f_{5/2}$-$\nu2p_{1/2}$ separation establishes the $N=34$ shell closure.}
\label{Fig:SFig4MG}
\end{figure*}